\documentclass[aps,prl,twocolumn,amsmath,amssymb,superscriptaddress,reprint]{revtex4-1}
\usepackage{amsmath}
\usepackage{graphicx}
\usepackage[dvipsnames]{color}
\usepackage{hyperref}
\usepackage{psfrag}
\usepackage{stmaryrd}
\usepackage{amssymb}
\usepackage{wasysym}
\usepackage{float}
\usepackage{color}

\usepackage[latin1]{inputenc}
\usepackage[T1]{fontenc}
\usepackage[french,english]{babel}
\usepackage{lmodern}

\begin{document}

\title{Shortcut to geostrophy in wave-driven rotating turbulence: the quartetic instability}

\author{Maxime Brunet}
\affiliation{Université Paris-Saclay, CNRS, FAST, 91405, Orsay, France}
\author{Basile Gallet}
\affiliation{Université Paris-Saclay, CNRS, CEA, Service de Physique de  
l'\'Etat Condensé, 91191, Gif-sur-Yvette, France}
\author{Pierre-Philippe~Cortet}
\email[]{ppcortet@fast.u-psud.fr} \affiliation{Université Paris-Saclay, CNRS, FAST, 91405, Orsay, France}

\date{\today}

\begin{abstract}
We report on laboratory experiments of wave-driven rotating
turbulence. A set of wavemakers produces inertial-wave beams that
interact nonlinearly in the central region of a water tank mounted
on a rotating platform. The forcing thus injects energy into
inertial waves only. For moderate forcing amplitude, part of the
energy of the forced inertial waves is transferred to subharmonic
waves, through a standard triadic resonance instability. This
first step is broadly in line with the theory of weak turbulence.
Surprisingly however,  stronger forcing does not lead to an
inertial-wave turbulence regime. Instead, most of the kinetic
energy condenses into a vertically invariant geostrophic flow,
even though the latter is unforced. We show that resonant quartets
of inertial waves can trigger an instability -- the ``quartetic
instability'' -- that leads to such spontaneous emergence of
geostrophy. In the present experiment, this instability sets in as
a secondary instability of the classical triadic instability.
\end{abstract}

\maketitle

\textit{Introduction.---} Turbulence subject to global
rotation~\cite{Cambon2008,Davidson2013,Pedlosky1987} can take
multiple forms depending on the large-scale mechanism driving it.
At the extremes of the spectrum are the Taylor-Proudman
theorem~\cite{Greenspan1968}, which states that rapidly rotating
flows tend to become two-dimensional (2D, or ``geostrophic'', invariant along the
rotation axis) and weak turbulence theory
(WTT)~\cite{Nazarenko2011}, which describes the flow as a
collection of nonlinearly interacting inertial waves. Both
 approaches address the low-Rossby-number
limit, but they are somewhat blind to one another: the Taylor-Proudman theorem is established for a slowly
evolving flow, thus ruling out inertial waves at the outset. On
the other hand, inertial-wave turbulence assumes that there is no
geostrophic flow to begin
with. The flow is then {\it protected} from the emergence of
geostrophy, because triadic interactions of inertial waves cannot
transfer energy from the fast wave modes to the slow geostrophic
one~\cite{Smith99,Cambon2008}: at this order of the
low-Rossby-number asymptotic expansion, the geostrophic component
 remains zero.

A pre-requisite to testing WTT in the laboratory is to be able to
generate an ensemble of nonlinear inertial waves. Because this is
an arduous task, most rotating turbulence experiments use instead
standard forcing devices of non-rotating turbulence:
grids~\cite{Hopfinger1982,Jacquin1990,Staplehurst2008,Lamriben2011},
jets~\cite{Baroud2003,Yarom2013,Yarom2014,Yarom2017}, vortex
generators~\cite{Campagne2014,Campagne2015},
impellers~\cite{Campagne2016}. These devices are rather
inefficient at forcing inertial waves, either because they lack
the right time-dependence to match the wave dispersion relation
(most of them are steady), or because they are directly compatible
with a Taylor-Proudman-like 2D flow. The
geostrophic component then dominates the flow in the low-Rossby-number
limit, with a small energy fraction in the
wave modes~\cite{Yarom2014,Campagne2015,Yarom2017,Gallet2015}.

To investigate nonlinear wave dynamics, there is thus a crucial
need for experiments with energy input into the wave modes only.
Experimental methods have arisen to generate isolated inertial-wave beams in
rotating
flows~\cite{Manders2003,Messio2008,Bordes2012,Duran-Matute2013,Machicoane2015,Brunet2019}, but the fate of a chaotic ensemble of inertial
waves remains unexplored in the laboratory. In this Letter, we
report on a rotating turbulence experiment where energy is input
into the wave modes only, with the following questions in mind:
what happens beyond the first triadic subharmonic instability? Is
inertial-wave turbulence really {\it protected} from the emergence
of a strong geostrophic mode? If not, what is the physical
mechanism responsible for the emergence of geostrophy?

\begin{figure}
    \centerline{\includegraphics[width=7cm]{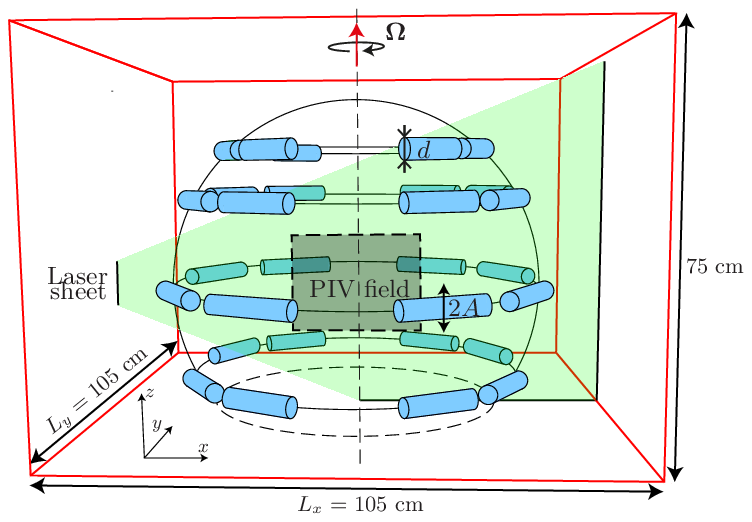}}
\caption{Experimental setup. 32 horizontal cylinders oscillate
vertically inside a tank centered on a platform rotating at
$\Omega=18$~rpm.}\label{fig:setup}
\end{figure}

\textit{Experimental setup.---} The experimental setup is sketched
in Fig.~\ref{fig:setup}. $32$ horizontal cylinders of diameter
$d=4$~cm oscillate vertically inside a parallelepipedic water tank
of $105\times 105$~cm$^2$ base and $75$~cm height, with an
amplitude $A$, an angular frequency $\omega_0$, and independent
random initial phases. The cylinders are tangent to an
80-cm-diameter  virtual sphere horizontally centered in the water
tank.

The apparatus is mounted on a two-meter-diameter platform rotating
at a rate $\Omega=18$~rpm around the vertical axis $z$. The cylinders oscillate at $\omega_0 = 0.84 \times
2\Omega$, each cylinder generating four inertial-wave
beams~\cite{Machicoane2015,Cortet2010} making angles $\pm
\arccos(\omega_0/2\Omega) \simeq \pm 32.9^\circ$ with the horizontal~\cite{Greenspan1968}. In the central region of the
sphere, these beams interact nonlinearly, producing a homogeneous
turbulent flow. The amplitude of oscillation $A$ is
varied from $1.0$ to~$8.2$~mm, which corresponds to a
forcing Reynolds number $130 \leq Re=A\omega_0 d/\nu \leq 1040$ and a forcing Rossby number $0.022\leq Ro=A\omega_0/2\Omega d \leq 0.172$. We access the velocity
field inside a vertical plane containing the center of the sphere
with an on-board Particle Image Velocimetry (PIV) system. Two velocity
components $(u_x,u_z)$ are measured with a
resolution of $1.94$~mm over an area $\Delta x \times \Delta z = 285
\times 215$~mm$^2$, twelve times per wavemaker period $T=2\pi/\omega_0$.

\begin{figure}
    \centerline{\includegraphics[width=8.3cm]{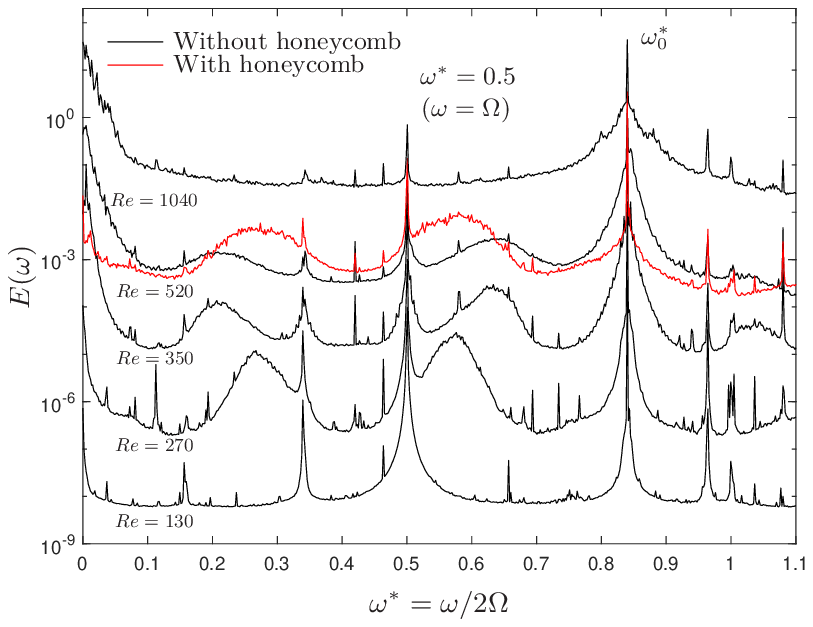}}
    \caption{Temporal spectrum of the velocity field. A vertical shift by a factor of $10$
    has been introduced between successive spectra. The red spectrum
    corresponds to an additional experiment at $Re = 520$ with a
    honeycomb bottom boundary.}\label{fig:spectrefreq}
\end{figure}

\begin{figure}
      \centerline{\includegraphics[width=8cm]{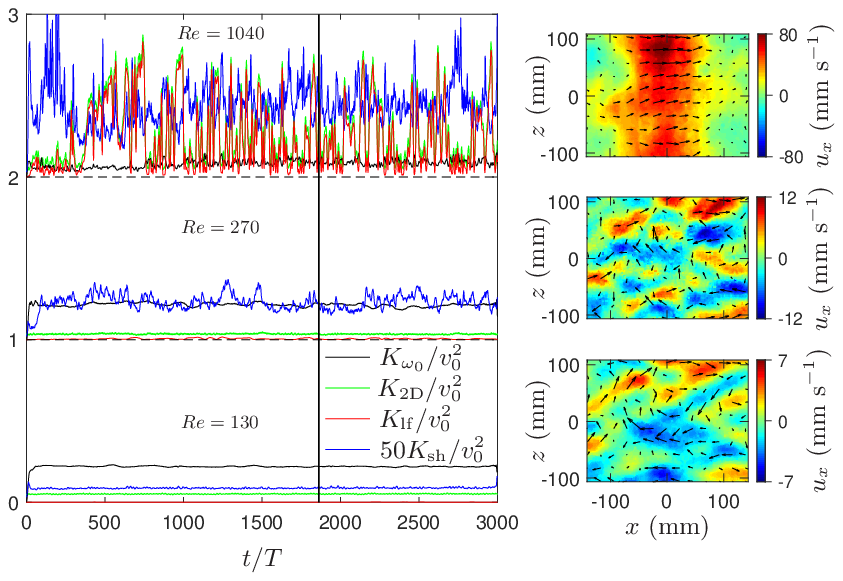}}
\caption{Time series of the kinetic energies $K_{\omega_0}$ of the
forced mode, $K_{\rm 2D}$ of the 2D mode, $K_{\rm lf}$ of the
low-frequency mode, and $K_{\rm sh}$ of the first subharmonic mode
(the latter being multiplied by $50$) for $Re=130;
270$ and $1040$ and normalized by $v_0^2=(A\omega_0)^2$. The
forcing starts at $t=0$. A vertical shift of $1$ has been
introduced between the curves at different $Re$. For each $Re$,
the right panel shows the corresponding velocity field at time
$t=1830\,T$.}\label{fig:ECvst}
\end{figure}

\textit{Triadic resonance instability.---} We show in
Fig.~\ref{fig:spectrefreq} the temporal energy spectrum of the velocity
field in statistically steady state. For the lowest amplitude
($Re=130$), the spectrum displays two dominant peaks, the next subdominant 
peaks being less energetic by two orders of magnitude. A first peak
at normalized frequency $\omega_0^*=\omega_0/2\Omega=0.84$
corresponds to the waves forced by the wavemakers, while a second
one at $\omega^*=\omega/2\Omega=0.5$ corresponds to a
``precession flow'' induced by the Earth
rotation~\cite{Boisson2012,Triana2012}. This spectrum illustrates the base-flow 
at low forcing amplitude: an ensemble of wide
inertial wave beams generated by the wavemakers and propagating
towards the central region (see the velocity fields in
Fig.~\ref{fig:ECvst}).

Increasing the forcing amplitude to $Re=270$, two subharmonic bumps
emerge around frequencies
$\omega_1^*=0.26\pm 0.04$ and $\omega_2^*=0.58\pm 0.04$. They
correspond to inertial waves created by the triadic resonance
instability of the primary waves, following the classical
scenario reported in various
experimental~\cite{Bordes2012,Joubaud2012,Scolan2013,Brunet2019}
and numerical~\cite{Koudella2006,Jouve2014,LeReun2017} studies.
The secondary frequencies notably satisfy a resonance condition
with the primary waves: $\omega_1^* + \omega_2^* \simeq
\omega_0^*=0.84$~\cite{footnote1}.

\textit{Emergence of geostrophy.---} In parallel, we observe the
emergence of a strong peak at $\omega^*=0$ in the high-$Re$
spectra. Low-pass filtering the velocity field with $\omega^* <
0.10$ reveals that this peak is associated with 2D geostrophic
vortices wandering in the horizontal plane. The snapshots in
Fig.~\ref{fig:ECvst} highlight this emergence of geostrophy.

In Fig.~\ref{fig:ECvst}, we show time series of the kinetic energies
$K_{\rm 2D}(t)$ in the vertically averaged velocity field inside the PIV 
domain, $K_{\rm lf}(t)$ in the low-frequency mode (using a low-pass
filter with $\omega^*<0.1$), $K_{\rm sh}(t)$ in the
first subharmonic bump produced by the triadic instability
(band-pass filter with $0.1<\omega^*<\omega_0^*/2$) and $K_{\omega_0}(t)$ in the forced waves (band-pass filter
with $|\omega^*-\omega_0^*|<0.1$). The time series of $K_{\rm 2D}$
and $K_{\rm lf}$ are almost identical, the difference being due to
some pollution of $K_{\rm 2D}$ by the large-scale
precession-induced flow at $\omega^*=0.5$. This confirms that the
low-frequency spectrum corresponds to the 2D
geostrophic flow. Most interestingly, Fig.~\ref{fig:ECvst}
highlights the bifurcation sequence by which the geostrophic flow
arises:
\begin{itemize}
\item For $Re=130$, the flow consists mostly in primary waves,
with very low levels of $K_{\rm lf}$ and $K_{\rm sh}$. \item For
$Re=270$, the system is above the threshold of the triadic
instability, with significant subharmonic energy $K_{\rm sh}$ and
still negligible energy $K_{\rm lf}$ in the low-frequency
geostrophic component. \item For $Re=1040$, the triadic
instability first sets in, inducing significant levels of $K_{\rm
sh}$, before the geostrophic mode arises and settles in a strongly
fluctuating state with large $K_{\rm 2D}\simeq K_{\rm lf}$~\cite{footnote2}.
\end{itemize}
The geostrophic flow thus arises above a
threshold value of the driving intensity, as a secondary
instability of the triadic instability. While the first triadic
instability is in line with the phenomenology of wave turbulence,
the system then rapidly finds a shortcut to directly transfer
energy to the geostrophic mode.

\begin{figure}
    \centerline{\includegraphics[width=8.3cm]{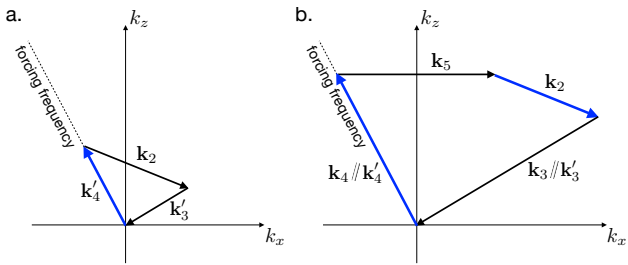}}
    \caption{An illustrative example of the quartetic secondary instability. 
    \textbf{a.} The triadic
    instability transfers energy from the mode ${\bf k}'_4$ (at the
    forcing frequency) to modes at ${\bf k}_2$ and ${\bf k}'_3$.
    \textbf{b.} One can build a resonant quartet by keeping ${\bf
    k}_2$, inserting a horizontal wavenumber ${\bf k}_5$, and closing
    the quartet with two wavevectors ${\bf k}_3$ and ${\bf k}_4$
    parallel to ${\bf k}'_3$ and ${\bf k}'_4$ respectively. ${\bf
    k}_4$ is energized by the forcing, while ${\bf k}_2$ has been
    energized at step a. Through a resonant quartet instability, the
    geostrophic mode ${\bf k}_5$ then spontaneously emerges, together
    with ${\bf k}_3$.}\label{fig:quartet}
\end{figure}

\textit{Quartetic secondary instability.---} We propose a generic
scenario to explain this shortcut to geostrophy: while resonant
triads cannot transfer energy to the slow geostrophic modes,
resonant quartets can, as discussed by Smith and
Waleffe~\cite{Smith99} (SW in the following). More surprisingly,
we will show that these resonant quartets can spontaneously
generate geostrophic flows through a ``quartetic instability'',
the four-wave equivalent of the triadic instability. To illustrate this 
mechanism, we focus on a single resonant triad, and an associated single 
resonant quartet. They are chosen somewhat 
arbitrarily, with the goal of providing a simple example of quartetic shortcut 
to geostrophy.

We thus consider the resonant triad in
Fig.~\ref{fig:quartet}a, with wavevectors ${\bf k}_2  = k_0
[6;0;-3]$, ${\bf k}'_3 = k_0 [-18/5;0;-9/5]$ and ${\bf k}'_4 = k_0
[-12/5;0;24/5]$ and polarities $(s_2=+1,s'_3=+1,s'_4=+1)$ in
the conventions of SW (see Supplemental Material). $k_0$ is an
arbitrary inverse length-scale that does not enter the analysis.
In a similar fashion to our experiments, we consider a system
where wavemakers generate waves at frequency $\omega({\bf k}'_4)$,
energizing wavevector ${\bf k}'_4$ but also many other parallel
wavevectors associated with the same frequency (see 
Fig.~\ref{fig:quartet}a)~\cite{footnote3}. Above a threshold amplitude of the 
forcing,
the system spontaneously transfers some energy from the mode ${\bf
k}'_4$ to ${\bf k}'_3$ and ${\bf k}_2$ through the triadic
resonance instability mechanism~\cite{Koudella2006,Bordes2012}.
Consider then the quartet sketched in Fig.~\ref{fig:quartet}b,
built by keeping wavevector ${\bf k}_2$, inserting a horizontal
wavevector ${\bf k}_5=k_0 [8;0;0]$, and closing the quartet with
two wavevectors ${\bf k}_3=k_0 [-10;0;-5]$ and ${\bf k}_4=k_0
[-4;0;8]$ parallel to ${\bf k}'_3$ and ${\bf k}'_4$ respectively.
The quartet $({\bf k}_2,{\bf k}_3,{\bf k}_4,{\bf k}_5)$ happens to be the one 
considered by Smith \& Waleffe~\cite{Smith99}. This quartet is
resonant for the mode polarities
$(s_2,s_3,s_4,s_5)=(+1,+1,+1,-1)$, which we adopt in the
following. The mode ${\bf k}_4$ is at
frequency $\omega({\bf k}_4)=\omega({\bf k}'_4)$ and is also
energized by the forcing, while the mode ${\bf k}_2$ has been
energized by the triadic instability at the previous step. We 
now show that the mode ${\bf k}_3$ and the geostrophic mode ${\bf
k}_5$ can emerge spontaneously through a quartetic instability.
Denoting as $b_i$ the amplitude of the mode of wavevector ${\bf
k}_i$ with the polarity $s_i$, a multiple-timescale expansion in
the low-Rossby-number limit yields the following quartetic
evolution equations:
\begin{eqnarray}
(\partial_t + \nu k_3^2) \, b_3 & = & C_{{\bf k}_{3};{\bf k}_{2};{\bf
k}_{4};{\bf k}_{5}}^{+1,+1,+1,-1} \, \overline{b_2}\, \overline{b_4}\,
\overline{b_5} \, ,
\label{eqb3}\\
(\partial_t + \mu + \nu k_5^2) \,  b_5 & = & C_{{\bf k}_{5};{\bf k}_{2};{\bf
k}_{3};{\bf k}_{4}}^{-1,+1,+1,+1} \, \overline{b_2}\, \overline{b_3}\,
\overline{b_4} \, , \label{eqb5}
\end{eqnarray}
where an overbar denotes the complex conjugate and the 
$\mu$-term mimicks bottom-drag acting on the geostrophic mode. The
details of the asymptotic expansion are deferred to the
Supplemental Material (SM), as well as the expressions of the
($\mu$- and $\nu$-independent) quartetic coefficients
$C_{{\bf k}_{3};{\bf k}_{2};{\bf k}_{4};{\bf
k}_{5}}^{+1,+1,+1,-1}$ and $C_{{\bf k}_{5};{\bf k}_{2};{\bf
k}_{3};{\bf k}_{4}}^{-1,+1,+1,+1}$. Our base state
consists in time-independent finite amplitudes $b_2$ and $b_4$,
energized respectively by the forcing and by the triadic
instability. We consider infinitesimal perturbations of the modes 3 and 5,
$b_3(t) \ll 1$ and $b_5(t) \ll 1$. Upon looking for perturbations
evolving as $e^{\gamma t}$ and neglecting the dissipative terms,
Eqs.~(\ref{eqb3}) and (\ref{eqb5}) yield:
\begin{eqnarray}
\gamma=\pm \sqrt{ C_{ \, {\bf k}_{3};{\bf k}_{2};{\bf k}_{4};{\bf k}_{5}}^{ \, +1,+1,+1,-1} \times \overline{C}_{ \, {\bf k}_{5};{\bf k}_{2};{\bf k}_{3};{\bf k}_{4}}^{ \,  -1,+1,+1,+1} } \,  |b_2|  |b_4| \, . \label{gammainviscid}
\end{eqnarray}
Because $C_{ \, {\bf k}_{3};{\bf k}_{2};{\bf k}_{4};{\bf k}_{5}}^{ \, +1,+1,+1,-1} \times \overline{C}_{ \, {\bf k}_{5};{\bf k}_{2};{\bf k}_{3};{\bf k}_{4}}^{ \,  -1,+1,+1,+1} >0$ (see SM), waves at ${\bf k}_3$ and ${\bf k}_5$ grow exponentially: a
geostrophic flow emerges spontaneously through this quartetic
instability. When the damping terms are retained,  the
amplification gain (\ref{gammainviscid}) must overcome viscous and
frictional damping. This occurs above a threshold value of the
product $|b_{2}|  |b_{4}|$, which we determine by performing a
similar linear stability analysis on the full equations
(\ref{eqb3}) and (\ref{eqb5}), before setting $\gamma$ to zero.
This yields the instability criterion:
\begin{eqnarray}
|b_2| |b_4| \geq \sqrt{\frac{\nu k_3^2 \, (\mu + \nu k_5^2)}{ C_{ \, {\bf k}_{3};{\bf k}_{2};{\bf k}_{4};{\bf k}_{5}}^{ \, +1,+1,+1,-1} \times \overline{C}_{ \, {\bf k}_{5};{\bf k}_{2};{\bf k}_{3};{\bf k}_{4}}^{ \, -1,+1,+1,+1} }} \label{threshold}\, .
\end{eqnarray}
In other words, the system needs to be sufficiently above the
threshold of the triadic instability, with large enough $|b_{2}|$,
to trigger the quartetic secondary instability.

\begin{figure}
    \centerline{\includegraphics[width=7.5cm]{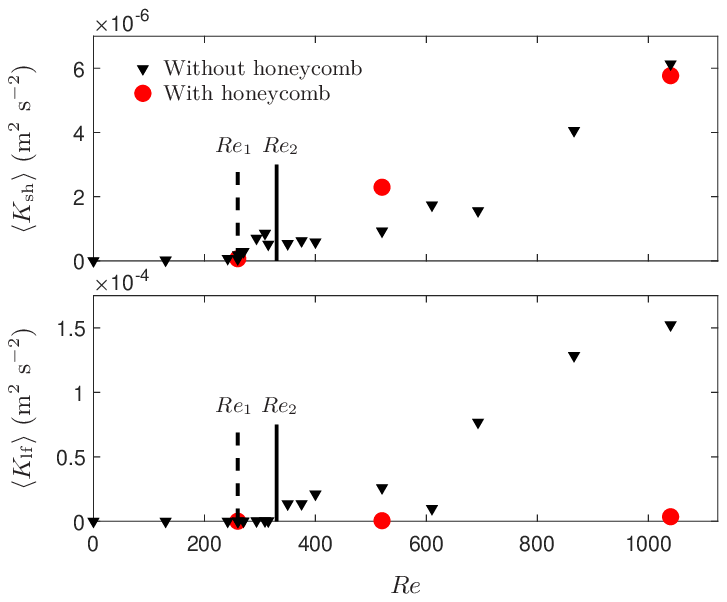}}
    \caption{$\langle K_{\rm sh}\rangle$ and $\langle K_{\rm lf}
    \rangle$ as functions of $Re$. The threshold Reynolds number
    $Re_2$ for the emergence of geostrophy is $25\%$ higher than the
    threshold $Re_1$ of the triadic instability when the bottom
    boundary is smooth. $Re_2$ is shifted to prohibitively large
    values when the bottom boundary is made rough using a honeycomb
    structure. The error bars on $\langle K_{\rm sh}\rangle$ are comparable to 
    the size of the symbols. The error bars on $\langle K_{\rm lf}
    \rangle$ are comparable to the size of the symbols for $Re\leq 600$, and 
    around $\pm 15\%$ for $Re > 600$.} \label{fig:bifurcation}
\end{figure}

To confront this scenario to the experimental data, we plot in
Fig.~\ref{fig:bifurcation} the time-averaged kinetic energies
$\left< K_{\rm sh} \right>$ and $\left< K_{\rm 2D} \right>$ as
functions of $Re$. The primary triadic instability
arises above a threshold value $Re_1 \simeq 260$, where $\left<
K_{\rm sh} \right>$ departs from zero, while the threshold of the
quartetic instability is $Re_2 \simeq 330$, where $\left< K_{\rm
2D} \right>$ departs from zero. One may notice that $\left< K_{\rm
2D} \right>$ has a non-monotonic behaviour at higher $Re$. This
feature goes beyond the scope of the present model, which
describes the immediate vicinity of the bifurcation points only. An interesting extension of the model could be to focus on scale separation between wave motion and large-scale geostrophic flow, to capture the four-wave instability through mean-field coefficients \cite{Frisch1987,Dubrulle,Sivashinsky}, in the spirit of Refs.~\cite{Kleeorin,Rogachevskii}.

Despite its simplicity ---a triad inside a vertical plane, and an
associated quartet inside that same plane--- the model
successfully captures the sequence of bifurcations observed in the
experiment: a primary triadic instability followed by a secondary quartetic instability, which
triggers the emergence of geostrophy. The frequency of the forced waves in the 
theoretical model being close to the frequency of the forced waves in the 
laboratory experiment (respectively $\omega({\bf k}_4)=\omega({\bf 
k}'_4)=0.89\times 2 \Omega$ and $\omega_0=0.84\times 2 \Omega$), we can compare 
quantitatively the various terms arising in the model to their experimental 
counterparts. We estimate both sides
of (\ref{threshold}) at the experimental threshold $Re_2$ where the geostrophy emerges: the amplitude of
the primary waves $|b_4|$ is approximately $\sqrt{\left<
K_{\omega_0}\right>} \simeq 4.5$~mm~s$^{-1}$. The square-root of
$\left<K_{\rm sh}\right>$ gives $|b_2| \simeq 0.7$~mm~s$^{-1}$,
and the left-hand side of (\ref{threshold}) is around $3.2\times 10^{-6}$~m$^2$~s$^{-2}$. The Ekman friction term is
here comparable to the viscous one, and the wavevectors have comparable lengths. We thus estimate the numerator of the
right-hand side of (\ref{threshold}) as simply $\nu k_4^2$.
Substituting $k_4=\sqrt{80} \, k_0$ and $C_{{\bf k}_3;{\bf
k}_{2};{\bf k}_4;{\bf k}_5}^{+1,+1,+1,-1} \times
\overline{C}_{{\bf k}_{5};{\bf k}_{2};{\bf k}_{3};{\bf
k}_{4}}^{-1,+1,+1,+1} \simeq 726 \, {k_0^4}/{\Omega^2}$ (computed
in the SM), the right-hand side of (\ref{threshold}) is estimated
as: $80 \, \nu \Omega /\sqrt{726} \simeq 5.6 \times
10^{-6}$~m$^2$~s$^{-2}$. We conclude that the two sides of
(\ref{threshold}) have comparable magnitudes at $Re_2$, which
confirms the expected balance between quartetic amplification and damping terms 
at threshold. 

Beyond these order of magnitude estimates, fully quantitative predictions would 
require an exact knowledge of the base flow, together with an optimization of 
the bifurcation threshold over all possible combinations of resonant triads and 
associated resonant quartets, a challenging task in general. Nevertheless, we 
can test the qualitative predictions of the theory further: we performed three 
complementary experiments with a honeycomb structure at the bottom of the tank
(mesh=1.7~cm, height=2~cm). We first sent an inertial-wave beam
towards this rough bottom boundary, and observed that it reflects
with negligible losses: the honeycomb structure has little effect
on wave-like motion. By contrast, we expect it to induce a strong
damping of the geostrophic modes, through a drastic increase in
bottom drag. In some sense, this honeycomb structure is an
experimental means of achieving the specific damping of the
geostrophic modes included numerically by Le Reun \textit{et
al.}~\cite{LeReun2017}. In line with
these expectations, we observe in Fig.~\ref{fig:bifurcation} that
the onset of the triadic instability remains unaffected, while the
threshold of the secondary quartetic instability is shifted to
prohibitively large values of $Re$~\cite{footnote4}. These observations are qualitatively
captured by expression (\ref{threshold}) at large $\mu$, this
large $\mu$ being a rough modelization of the possibly turbulent
and quadratic bottom drag induced by the honeycomb structure. This
provides further experimental evidence that the subharmonic waves
and the geostrophic flow appear through distinct instabilities.

\textit{Discussion.---} Although our study focuses on rotating
turbulence, the phenomenology probably holds for the various
anisotropic fluid systems that can be decomposed into fast wave
modes and a slow manifold: Rossby waves and zonal flows on the
beta plane, internal gravity waves and shear flows in stratified
fluids, etc. As a matter of fact, the ability of quartets of
inertial waves to transfer energy to the geostrophic modes was
soon identified by SW, based on previous work by
Newell~\cite{Newell69} showing that quartets of Rossby waves can
force zonal flows. These studies focus on the direct forcing of
the slow manifold by three pre-existing waves. By contrast, here
we force waves at a single frequency. Even in that case, however,
our experimental results show that the system efficiently
transfers energy to the 2D flow through a quartetic instability. 
A somewhat similar situation was recently 
reported by Le~Reun \textit{et al.}~\cite{LeReun2019}. In their experiment, the 
libration of an ellipsoid induces a 2D base-flow that is unstable to 
inertial-wave motion. As one increases the driving amplitude, the system first 
transfers energy to secondary inertial-wave modes through triadic resonances, 
before inducing a 
geostrophic vortex on top of the base librating flow. The authors conclude that 
this geostrophic vortex probably emerges because the Rossby number is only 
moderately low, although they also invoke the possible role of an instability 
described by Kerswell in cylindrical geometry \cite{Kerswell1999}. Although it 
was not identified as such at the time, we believe that Kerswell's instability 
may very well be underpinned by the present four-wave mechanism: the 
corresponding quartet would consist of two wave modes and two geostrophic 
ones,  suggesting a scenario where geostrophy arises through a direct quartetic 
instability (as opposed to a secondary one). While it is difficult to 
discriminate between the direct and secondary quartetic instability scenarii 
experimentally, we expect the direct instability to be more strongly stabilized 
by bottom friction than the secondary instability, because it involves two 
geostrophic modes. By contrast, the secondary instability involves a single 
geostrophic mode, hence a single factor $\sqrt{\mu}$ on the right-hand side of 
(\ref{threshold}).

The quartetic instability induces a transfer of energy to the slow
modes, thus restricting the domain of validity of WTT. Indeed, the
relation~(\ref{threshold}) indicates that this instability sets in
when the Ekman number is reduced at fixed Rossby number. The only
hope to observe weak wave turbulence is then to focus on
distinguished limits where the Rossby number goes to zero faster
than some power of the Ekman number~\cite{LeReun2017}, the precise
boundary in parameter space depending on the dominant damping
mechanism (bulk viscosity, Ekman friction, etc.). For wave
turbulence to develop, this damping mechanism must be dominant at
the quartetic order, but negligible at the triadic one. The
corresponding parameter range may be too narrow for an
experimental study. A more promising approach could be the use of
rough boundaries, such as the honeycomb structure we placed at the
bottom of the tank, to induce turbulent Ekman layers and
preferential damping of the geostrophic flow.

\acknowledgments We acknowledge J. Amarni, A. Aubertin, L. Auffray
and R. Pidoux for experimental help. BG thanks T. Le Reun, B.
Favier and M. Le Bars for insightful discussions. This work has
been supported by the Agence Nationale de la Recherche through
Grant ``DisET'' No.~ANR-17-CE30-0003. BG acknowledges support by
the European Research Council under grant agreement 757239.


\begin{thebibliography}{99}
\bibitem{footnote2}
These fluctuations are due to the geostrophic vortices coming in and out
of the PIV plane.
\bibitem{footnote4}
See the associated spectrum in Fig.~\ref{fig:spectrefreq}, where the 
zero-frequency geostrophic
peak has disappeared.
\end{thebibliography}
\end{document}